# Adaptive Finite Differencing in High Accuracy Electronic Structure Calculations

E.L. Briggs,[1,*] Wenchang Lu,[1,2,*] and J. Bernholc[1,2,*]

1. Department of Physics, North Carolina State University, Raleigh, NC 27695-8202

2. Computational Sciences & Engineering Division, Oak Ridge National Laboratory, TN 37831

* E-mails: elbriggs@ncsu.edu, wlu2@ncsu.edu, bernholc@ncsu.edu

A new method for implementing the kinetic energy operator for real-space, grid-based electronic structure codes is developed. It is based on multi-order Adaptive Finite Differencing (AFD) and uses atomic pseudo orbitals produced by the corresponding pseudopotential codes to optimize the standard finite difference (SFD) operators for improved accuracy. Results are presented for a variety of test systems and Bravais lattice types, including the well-known Δ test for 71 elements in the periodic table, the Mott insulator NiO, and borax decahydrate, which contains covalent, ionic, and hydrogen bonds. The tests show that an $8^{th}$-order AFD operator leads to the same average Δ value as that achieved by plane-wave codes and is typically far more accurate and a much lower computational cost than a $12^{th}$-order SFD operator. The scalability of real-space electronic calculations is demonstrated for a 2,016-atom NiO cell, for which the computational time decreases nearly linearly when scaled from 18 to 144 CPU-GPU nodes.

# INTRODUCTION

Density functional theory (DFT) [1–3] has enabled accurate, *ab initio* predictions of materials properties and explanations of a wide range of experimental data. Tens of thousands of DFT calculations are performed each year. The reliability of DFT predictions has led to materials-genome-type projects,[4–7] in which a large set of possible material compositions and structures are screened by DFT calculations in order to identify those with promising properties. Those with the most potential are then suggested or selected for experimental synthesis and evaluation.

Reliability of predictions is critical to any evaluation of potential predictions, and the accuracy of DFT calculations is thus a factor when choosing a suitable computational method. However, full-precision DFT calculations can be computationally expensive, especially if complex structures or multi-component systems are involved, and the choice of approach and code often involves a tradeoff between accuracy and practicality. Fortunately, the methodology has advanced to the point where the results of several widely-used DFT codes agree well with each other and a set of benchmark high-accuracy calculations for a large set of elements across the periodic table.[8] However, the tradeoffs remain for more complex systems or large survey studies, requiring practical compromises between the computational expense, accuracy, and even the choice of the computational method. This paper describes a significant advance in the class of methods that use real-space grids to solve DFT equations. By using adaptive finite differencing to discretize the kinetic energy operator, the real-space results agree with those of plane-wave-based codes using much lower grid densities than those previously required and reproducing the benchmark DFT results[8] at the same accuracy level as those of plane-wave-based and all-electron codes. The improved discretization enables high-accuracy real-space calculations at a substantially reduced cost while leveraging the well-known advantages of real-space methods of easy parallelization across many nodes, including multi-CPU and multi-GPU configurations, and avoiding the use of Fast Fourier Transform algorithms, which require global communication across nodes.

Briefly, in density functional theory, the ground state electronic structure of many-electron systems is solved numerically, typically after introducing the Kohn-Sham equation,[2] which is derived variationally from the total energy functional after approximating the kinetic energy by that of non-interacting electrons. The electrons are described as single particles in the presence of an external potential stemming from nuclei and externally applied fields and an effective potential, introduced by interactions with other electrons and the effects of the Pauli exclusion principle and spin:

$$[-\nabla^2 + V_{ext}(r) + V_{eff}(r)]\psi_i(r) = \varepsilon_i \psi_i(r) \quad i = 1, \ldots N \quad (1)$$

for the *N* electrons in the system. The effective potential is further split into the classical electron-electron repulsion part and the exchange-correlation part that accounts for all effects left out by the classical non-interacting electrons picture. While it has been proven that the exact exchange-correlation potential is only a functional of the total electron density,[1] its functional form is unknown. Nevertheless, excellent approximations exist, starting with the local density approximation, which pointwise approximates this potential with that of the uniform-density electron gas through a "Jacobs ladder" of functionals with increasing complexity and accuracy[9]. Non-local approximations to exchange-correlation functionals have also been devised, including various forms of hybrid functionals.[10–12]

Depending on the treatment of the external potential from the nuclei, the methods to solve the Kohn-Sham equation can be categorized into two groups, the first being the all-electron method that includes both core and valence electrons in the calculations.[13–15] The other group replaces the nuclei and inert core electrons with atom-derived pseudopotentials[16–19] or effective core potentials[20–22] that are constructed to accurately describe the interactions of valence electrons with the ionic cores. This dramatically reduces the number of electrons that must be included in calculations. Furthermore, because the valence electrons experience a softer pseudopotential instead of the real potential, the wave functions are smoother, and the basis set required to describe them is much smaller. Another advantage is the implicit inclusion of relativistic effects in the pseudopotential so that relativistic effects do not need to be explicitly addressed in the numerical solution of the Kohn-Sham equations.[23] For these reasons, most large DFT calculations use pseudopotentials. When pseudopotentials are carefully constructed from atomic all-electron calculations, they are transferable to different atomic environments and provide results that agree well with those of all-electron calculations.[8,24]

The norm-conserving pseudopotentials[16,19] use angular momentum projectors to force pseudo wavefunctions to agree exactly with atomic all-electron wavefunctions beyond a matching radius, typically chosen slightly beyond the core radius. A Kleinman-Bylander form of pseudopotentials[25] converts the angular projectors to limited-range 3D projectors, significantly simplifying their applications in calculations that use plane-wave basis sets. Ultrasoft pseudopotentials[17] relax the norm-conservation criteria and further reduce the basis set size but introduce sharp augmented charge densities near the atomic cores that require careful numerical treatment. Recently, Hamann[19] incorporated the main aspects of ultrasoft pseudopotentials into a norm-conserving form and generated pseudopotentials with multiple projectors for each angular momentum. As a result, these pseudopotentials are nearly as soft as the ultrasoft ones while avoiding the complexity of the augmented charge density.

In the last few decades, considerable work has been devoted to developing methods for solving the Kohn-Sham equations using pseudopotentials. The wave functions are usually expanded in a basis set, and the resulting secular equations are solved using matrix methods, which provide wave function information used to analyze physical phenomena. Many physics-, chemistry- or mathematics-motivated bases exist, including plane waves[26–32], atomic or atomic-like orbitals,[33–36] Gaussians,[37–44] real-space grids,[45–60] wavelets,[61–64] and finite elements.[65–67]

The most widely used approaches use plane waves as a basis, taking advantage of the fact that the kinetic energy operator is diagonal in Fourier space, while the potential is diagonal in real space (except for the non-local projectors and non-local exchange). The transformations between the two spaces use Fast Fourier Transforms (FFTs), which have the well-known *MlogM* scaling with the number of plane waves or grid points. Another advantage of plane-wave calculations is using a single, easy-to-understand parameter determining the basis set size and thus the accuracy: the highest kinetic energy of plane waves included in the calculations – the plane-wave cutoff.

Since real space is the dual to Fourier space, methods using uniformly-spaced grids[45–49,52–60] should offer similar advantages to plane waves and avoid extensive use of FFTs, the price being an inexact treatment of the kinetic energy operator. As shown in the sections below, adaptive discretization of this operator leads to dramatically improved accuracy, facilitating grid-based calculations with modest grid sizes and enabling highly accurate large-scale calculations on modern parallel computers.

The rest of the paper is organized as follows. We first present an analysis of the finite difference operator in the context of electronic structure calculations, followed by a procedure to generate standard finite difference operators for non-orthogonal lattices. Next, we introduce an adaptive finite difference operator optimized for electronic structure calculations and present the results of the Δ test, a well-known test[8] evaluating the accuracy of electronic structure codes, comparing grid-based results to highly accurate all-electron benchmarks for 71 elements. The following subsections examine energy convergence with grid spacing, and show an example of a highly accurate, large-scale DFT calculation obtained using the adaptive operator, parallelized over the hundreds of CPU-GPU nodes of an exascale supercomputer. All the grid-based calculations use the Real-space MultiGrid (RMG) code.[47,48,68]

**RESULTS**

**Analysis of the finite difference operator**

Real-space, grid-based methods for solving the Kohn-Sham equations typically use finite differences (FD) to implement the kinetic energy operator $-\frac{1}{2}\nabla^2$ appearing in the Kohn-Sham equation. The space locality inherent to FD operators is computationally efficient and suitable for parallelization via domain decomposition, which partitions the computational volume into subvolumes assigned to different nodes. In plane-wave methods, the Laplacian operator is trivial, while the charge density calculations and applications of the potential need transferring of all the wave functions from the plane wave representation to real space using Fast Fourier Transforms (FFTs), which require communications across the entire computational domain. However, the plane-wave treatment of the Laplacian is exact within the basis set defined by the G-space cutoff, while the FD approximation introduces errors. Minimizing the FD error is thus necessary to exploit the computational advantages possible in real space. The various approaches have included non-uniform grids,[45] higher-order central finite difference operators,[46,69] the Mehrstellen discretization in our previous version of the RMG package,[47,48] and adaptive coordinates and grids.[50,70]

We begin the analysis by considering the standard central finite difference approximation for the second derivative of a function $f(x)$ on a uniform grid of spacing $h$, using $2n + 1$ grid points, as given by Eq. (2).

The coefficients $a_i$ are chosen to make the expression exact for polynomials of degree less than or equal to $2n$

$$f_n''(x_0) = a_0 f(x_0) + \sum_{i=1}^{n} a_i \big(f(x_0 + ih) + f(x_0 - ih)\big) + R_n(h^{2n}), \qquad (2)$$

where the truncation error is

$$R_n = b_n h^{2n} f^{2n+2}(x_0) + O(h^{2n+2}) \qquad (3)$$

and the $b_n$ arise from the Taylor series expansions used to derive Eq. (2), with $b_1 = \frac{1}{12}, b_2 = \frac{1}{24}$ etc. There are alternate representations of the Laplacian where the coefficients are optimized for a different set of properties. For example, controlling dispersion is required for stable finite difference solutions to the wave equation, and there is a large body of work on this specific topic[71–73]. The goal is to find coefficients for use in Eq. (2) that are optimized for electronic structure calculations. Let $f(x)$ be represented by a plane wave expansion of the form below

$$f(x) = \sum_{j=1,N} C_j \exp(iG_j x). \qquad (4)$$

The first term of the truncation error in Fourier space can be written as

$$R_n = b_n h^{2n} \sum_{j=1,N} (iG_j)^{2n+2} C_j \exp(iG_j x_0). \qquad (5)$$

For each frequency, the truncation errors of different orders are proportional to the $C_j$'s, which are determined by the Fourier transform of the original function. Therefore, a linear combination of finite difference operators of different orders can be constructed to reduce the truncation error. The new adaptive operator can be written as

$$f''_{\{new\}}(x_0) = (1+M)f''_n(x_0) - Mf''_{n-1}(x_0). \tag{6}$$

The new composite operator is guaranteed to be exact for polynomials up to degree $2(n-1)$, while a further variation of $M$ may be used to minimize the truncation error for a target class of functions. This method of optimizing finite difference operators by mixing stencils of a different order is not necessarily appropriate for arbitrary functions but has been used successfully for specific purposes.[73] In our RMG implementation and tests, we found that $M$ is always positive. This is supported by the error analysis below.

Analyzing the error in more detail is useful when the adaptive operator is applied to plane waves. For a given reciprocal vector G, the truncation errors for a standard n-th order operator can be expressed as a power series

$$R_n^s(G) = \frac{1}{h^2} \sum_{m=1,2\ldots} b_{n,m}(ihG)^{2n+2m} \tag{7}$$

Here we ignore the phase term $e^{iGx_0}$. For a periodic system, the values of $hG$ are always in the range of $[-\pi,\pi]$. The $b_{n,m}$ decrease with increasing $m$, and the signs of the errors alternate for m = even or odd, but the errors for hG $\gg$ 1 are still large compared to the small $hG$ range.

The truncation error for the adaptive $n$-th order operator is

$$R_n^a(G,M) = (1+M)R_n^s(G) - MR_{n-1}^s(G) \tag{8}$$

For a given $n$, the constants $b_{n,m}$ may be evaluated using a symbolic math package (e.g., SymPy[74]). For the 8th order adaptive operator ($n$ = 4) we obtain the truncation error

$$R_4^a(G,M) = \frac{1}{h^2}\left[-\frac{M}{560}(hG)^8 + \left[\frac{1}{3156} + \frac{M}{1680}\right](hG)^{10} - \left[\frac{1}{13860} + \frac{19M}{201600}\right](hG)^{12} + \cdots\right] \tag{9}$$

Fig. 1 is a graph of the truncation error vs. G from Eq. 9 for *M*=0.33 as well as graphs of the errors for the 6th and 8th order standard operators used to construct the adaptive. It's clear from the graph and Eq. 7 that the error increases rapidly as hG → 1. Therefore, a necessary condition for using finite differences in electronic structure calculations is for the wavefunction components to be small when |hG| >1 and approach zero as |hG| is close to its maximum π. Fig. 1(a) shows that for $G > X \cong 0.29$ the adaptive operator is more accurate than the standard 8th while the zoomed view in Fig. 1(b) shows that for $G < X$ the standard 8th order is slightly more accurate. The absolute magnitude of the errors is very different in the two regions, with the adaptive operator having a significantly flatter error curve over a broader range than the standard operators. The total truncation error depends on *both* $R_n^a(G, M)$ and the profile of $\psi(G)$, with the adaptive process optimizing *M* to minimize the total error. However, the calculations presented later show that even a fixed value of *M*=0.33 leads to more accurate results for a wide range of atomic species and grid spacings. While the reason for this is not immediately obvious, it can be understood by focusing on the differential absolute errors in the regions to the left and right of $G = X$ in Fig. 1. The absolute errors for plane waves with $G < X$ are small for both adaptive and standard operators. In a high-cutoff calculation, where the wavefunctions have little weight for $G > X$, the overall error and the differences between the operators are small. For a less converged calculation, where the wavefunctions have significant components for $G > X$, the adaptive operator has a significantly lower total error because the errors for the standard operators increase rapidly in this region. As noted earlier, the 8th-order adaptive operator with *M*=0.33 performs well over a wide range of atomic species and grid spacings, but as the grid density becomes very high, there is a point where the standard operators are more accurate because the wavefunctions have vanishing components for $G > X$. The fully adaptive operator overcomes this issue by optimizing *M* for specific calculations, decreasing *M* as the convergence level increases.

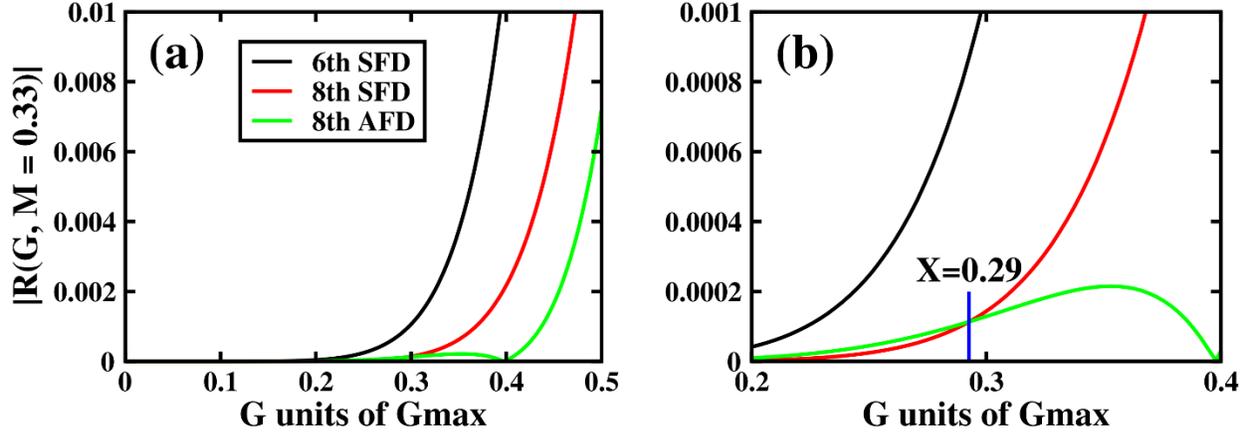

Fig. 1. Absolute value of the finite difference truncation error as a function of G, up to the Nyquist frequency ½ $G_{max}$. The $8^{th}$-order AFD operator uses a mixing value of *M*=0.33.

**Finite difference operator by Taylor Series expansion**

For cubic or orthorhombic lattice systems, the lattice vectors are orthogonal to each other, and the FD operators can be obtained separately for each direction. For a general lattice type, where the lattice vectors are not orthogonal, such as monoclinic or triclinic systems, the following process is used to generate the initial coefficients for a specific order. Then, the procedure discussed in the next subsection is used to optimize the FD operator further to reach high accuracy without increasing the computational cost.

First, a few special axes in the real-space grid are chosen appropriately for the lattice type; axes that coincide with the lattice vectors are always included. *2n* grid points along each axis are used for the following minimization. For each point, the function f(x,y,z) can be expressed as the Taylor series expansion

$$f(x_i, y_j, z_k) = f(x_0, y_0, z_0) + \sum_{lmp} \frac{f^{(lmp)}}{l!\,m!\,p!} h_x^l h_y^m h_z^p + \delta(ijk) \tag{10}$$

Where $f^{lmn}$ are the derivatives to be determined and $l+m+p < 2n+1$ for the 2n-order FD operator. We define an error norm $R$ by

$$R = \sum_{ijk} \frac{|\delta(ijk)|^2}{h_{ijk}^3} \tag{11}$$

where $\delta(ijk)$ is defined in equation (10) and $h_{ijk}$ is the distance from the point $(x_i, y_j, z_k)$ to the point $(x_0, y_0, z_0)$. The summation is over the set $\{ijk\}$ of points included in the FD operator. For monoclinic or triclinic systems, the sampling grid points are on axes along the lattice vectors and also additional axes. The FD coefficients can be obtained by minimizing the norm in Eq. 11, and we have

$$\frac{\partial R}{\partial f^{(lmp)}} = 0 \tag{12}$$

This will generate a set of linear equations and the derivatives $f^{(lmp)}$ can be obtained by solving these linear equations. [75] One can always avoid the singularity of the linear equations by including more grid points and axes.

For orthogonal lattices, the resulting FD kinetic energy operator is separable. For non-orthogonal lattices, cross terms occur, but both the kinetic energy operator and the first derivatives are obtained from the same sampling points with predetermined coefficients. Recently, a Kronecker product formulation of the kinetic energy operator for non-orthogonal lattices has been developed.[76] All of the derivatives are applied separately on the lattice vectors, but the cross terms $\partial^2/\partial\zeta\partial\eta$ need to be applied sequentially on each lattice vector. We have not explored this formulation.

**Adaptive finite difference operator**

A generalization from the one-dimensional mixed operator case discussed above to three dimensions is trivial, i.e., the truncation error in one lattice frequency is always proportional to the component of the target function at that frequency. This is true for all lattice types in solid-state physics, and the operators along the different axes are determined by the method described above.

For electronic structure applications in which the Kohn-Sham equation is solved numerically, the target functions are the Kohn-Sham wave functions. To a first-order approximation, these may be written as a linear combination of the pseudo atomic orbitals $\psi_{atomic}$. In a non-atomic environment, the $\psi_{atomic}$ should still provide a good representation of the Kohn-Sham wave functions in the regions near the ionic cores, as is implicitly assumed in the linear combination of atomic orbitals (LCAO) methods for calculating the electronic structure of molecules and solids.[77] In the interstitial regions, their deviation from the pseudo atomic orbitals may be larger, but as they are much less rapidly varying than in the core regions, an operator of the above form, which is exact for polynomials of degree $2(n-1)$, is highly accurate for a sufficiently large $n$. For these reasons, the pseudo atomic orbitals were chosen as the optimization functions for the finite difference operators. In analogy with pseudopotentials, operators constructed using this procedure are highly accurate for isolated atoms and should also be transferrable to solids or molecules.

The optimization procedure requires a reference standard. For a function discretized on a uniform real space grid, the most accurate Laplacian operator is the FFT method. Unfortunately, this would require transforming all the real-space wave functions to the reciprocal space via FFTs. For very large calculations in which the wave functions are distributed across many computing nodes, the FFTs are too slow because they require global communications.

Nevertheless, the FFT Laplacian is well-suited as a reference standard because it only needs to be computed once for each atomic orbital. The overall computational cost is negligible. We choose the objective function to be minimized as the norm of the difference between the kinetic energies of the real-space atomic orbitals computed via FFT versus their values computed by finite differences. Weighting factors are applied based on the atomic orbital occupations and the number of atoms of each kind in the anticipated large calculation. The final formula is

$$F(M) = \sum_{i=1,N} w_i |<\psi_i(r)|L_{FD}(M) - L_{FFT}|\psi_i>|^2, \qquad (13)$$

where $\psi_i$ is one of the distinct atomic orbitals, and $w_i$ is the weight factor equal to the total occupancy for this orbital. $L_{FD}$ is the finite differential Laplacian operator and $L_{FFT}$ is the FFT Laplacian operator. The minimization process for $F(M)$ can be performed by using a least-squares fit of the resulting data to determine the minima. If the deviation is too large for any orbital than a given tolerance, usually $10^{-4}$, it indicates that the grid spacing needs to be decreased. Alternatively, one can automatically determine the needed grid density by requiring that all deviations are below the tolerance.

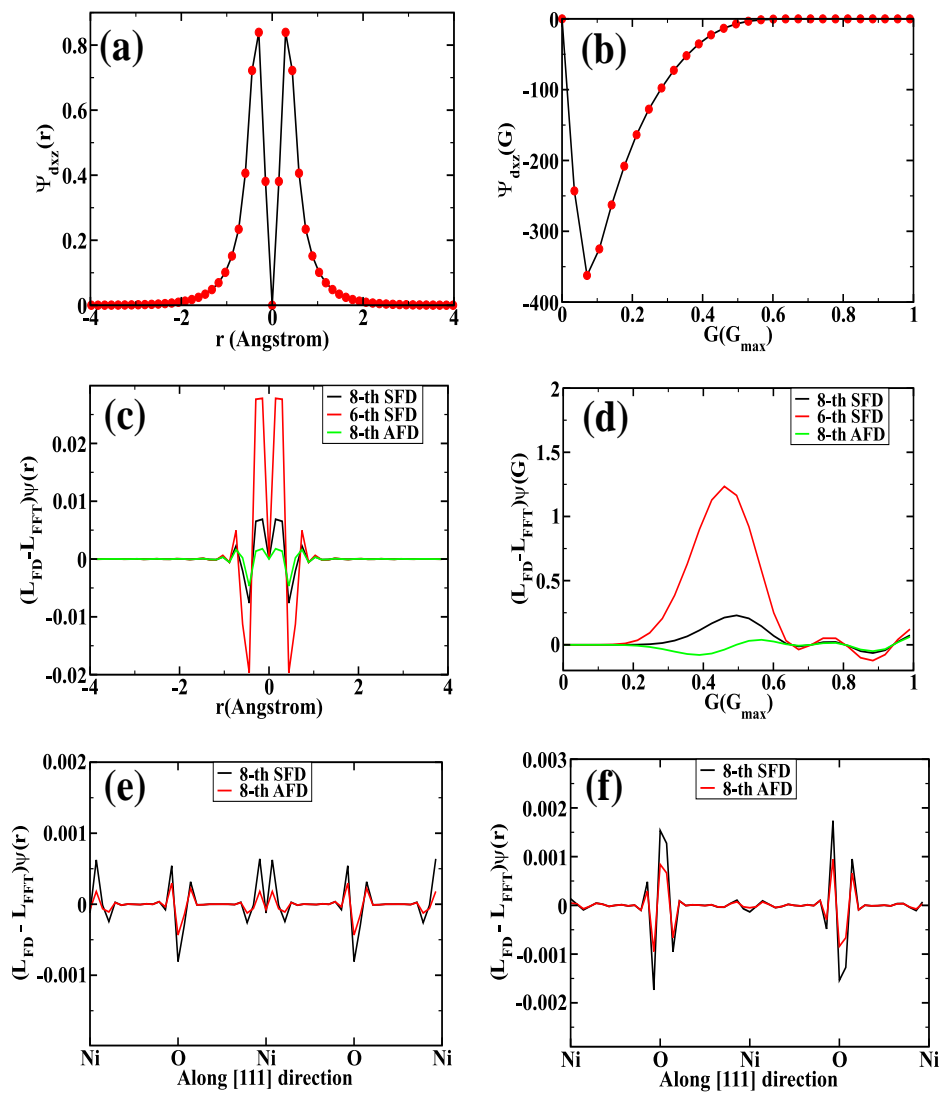

Fig. 2. Orbital and discretization errors in real space and reciprocal G space. Ni $3d_{xz}$ orbital in (a) real space and (b) G space. Discretization errors of Ni $3d_{xz}$ orbital in (c) real space and (d) G space. Finite difference errors for (e) HOMO and (f) LUMO wave functions of NiO along the [111] direction. Ni-O distance is 3.61 Å.

Figs. 2(a) and 2(b) show the Ni $3d_{xz}$ orbital in real and G spaces, respectively. Although the truncation errors for the high-frequency plane waves are large, the orbitals have very small components in this range, which can be clearly seen in Fig. 2(b). Figs. 2(c) and 2(d) show the truncation errors in real and G spaces for this orbital with different finite-difference operators. As was noted before, the truncation error at a specific G vector is proportional to the orbital's component at this frequency. This can be clearly seen in Fig. 2(d), where the errors have the same trends for both $6^{th}$- and $8^{th}$-order standard FD operators, independent of the grid spacings or the effective cutoff energies. In real space, the errors also have the same trends for both $6^{th}$- and $8^{th}$-order SFDs. Therefore, the errors are dramatically reduced with the adaptive $8^{th}$-order FD operator. In fact, it has better accuracy than the standard $12^{th}$-order FD operator (see below). The optimization uses the atomic orbitals as the target functions, and the resulting AFD is transferable to the wavefunctions in the real applications. Figs. 2(e) and 2(f) show the finite-difference errors (compared to the FFT results) for the highest occupied molecular orbital (HOMO) and the lowest unoccupied molecular orbital (LUMO) of NiO in a 64-atom cell. In the interstitial regions, the wavefunctions are smooth, and the errors from both SFD and AFD are small. In the regions near atoms, the errors from AFD are significantly smaller than those from SFD. This indicates that AFD is more accurate than SFD not only for the atomic orbitals but also for the extended-state wave functions. The more accurate total energy in the next section also proves that. Since a high-order FD operator requires more computational resources than the lower-order one, not only because of the size of the FD tensor but also the communication cost of reaching adjacent domains when domain decomposition is used in massively parallel calculations, the adaptive finite difference (AFD) operator leads to more accurate results at a significantly lower computational cost. We discuss the computational advantages of AFD below.

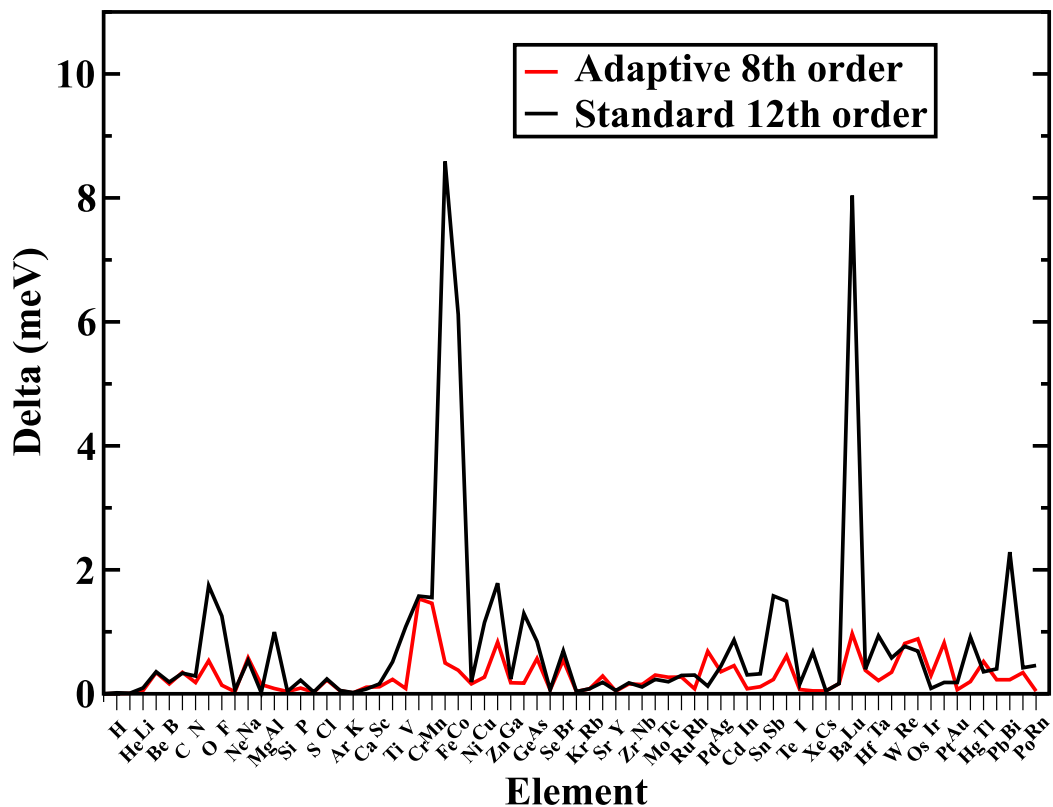

Fig. 3. Δ value comparison [meV] for individual atomic species calculated by RMG version 5.3 for the 8th order adaptive and 12th order standard operators.

**Accuracy across the periodic table**

The DFT Δ test[8,78] was selected to test the new operators in the RMG code[79] across a wide range of atomic species and lattice types. The Δ test evaluates the average variance between the equations of state calculated by different codes and has become a standard way of comparing the accuracy and reproducibility of DFT codes using different all-electron or pseudopotential methods, basis sets, and algorithms for 71 different atomic species. Published results listing the average Δ are available[80] for a large selection of DFT codes, using highly-converged all-electron results obtained by the WIEN2k code[81] as the reference standard. For our tests, we used a mixture of ONCV, GBRV, and pslibrary pseudopotentials which tracked closely with the SSSP accuracy library.[82] The correspondence was not exact, as RMG does not support PAW-based pseudopotentials used for some atomic species in the SSSP library. Fig. 3 provides an element-by-element comparison of the individual Δ values for the 8$^{th}$-order adaptive and the 12$^{th}$-order standard finite difference operators, while Table 1 lists the average Δ values for the 8$^{th}$, 10$^{th}$ and12$^{th}$- order standard operators as well as the 8$^{th}$-order adaptive and 8$^{th}$-, 10$^{th}$-and 12$^{th}$-order operators with fixed mixing. The latter are generated using the same mixing value for all species, making them easy to implement. They produce lower delta values than the standard operators of the same order, although not as low as the fully adaptive 8$^{th}$.

**Table 1**. Average delta values for various finite difference methods.

| Finite difference method | Δ value (meV) |
| --- | --- |
| 8$^{th}$ order standard | 3.753 |
| 10$^{th}$ order standard | 1.540 |
| 12$^{th}$ order standard | 0.802 |
| 8$^{th}$ order fixed (M=0.33) | 1.241 |
| 10$^{th}$ order fixed (M=0.416) | 0.553 |
| 12$^{th}$ order fixed (M=0.5) | 0.486 |
| 8$^{th}$ order adaptive | 0.298 |

**Energy convergence with the grid spacing**

In this subsection, we examine the convergence of the total energy versus grid resolution, represented by the corresponding plane-wave kinetic energy cutoff, of various forms of finite-differencing of the kinetic energy, compared to a reference value obtained by a plane-wave code that treats the kinetic energy exactly at that cutoff. For a given grid, we compute the kinetic energy cutoff as the radius squared of the inscribed sphere in the corresponding reciprocal space, which is consistent with the procedures used in plane wave codes. While the Δ test consists of calculations for elemental solids, we use supercells with more than one type of atom here, employing norm-conserving pseudopotentials (NCPP).[16,19] While the AFD method is equally applicable to ultrasoft pseudopotentials (USPP),[17] the wavefunctions from NCPP are typically more rapidly varying than those from USPP and therefore provide a more rigorous test of the quality of the operator. The reported results were generated using the SG15 potentials developed by Schlipf and Gygi [83] using Hamann's ONCV code.[19,84] The first system is the Mott insulator NiO, represented by a 64-atom supercell of NiO in an antiferromagnetic configuration, with Ni $3s^2 3p^6$ semi-core states included in valence. Fig. 4(a) shows the difference in total energy $\Delta E$ between a reference value and RMG v.5.3 using various forms of the kinetic energy operator. The reference values were generated using the open-source plane-wave code Quantum Espresso (QE)[30,85] that has an exact representation of the operator within its basis set. While all of the finite-difference-based operators achieve good agreement with QE at high energy cutoffs, the adaptive operators are considerably more accurate at lower cutoffs. Indeed, the 8$^{th}$-order AFD operator is more accurate than the 12$^{th}$-order standard operator up to a 205 Rydberg cutoff, with the difference at 124 Rydbergs being a factor of 11.

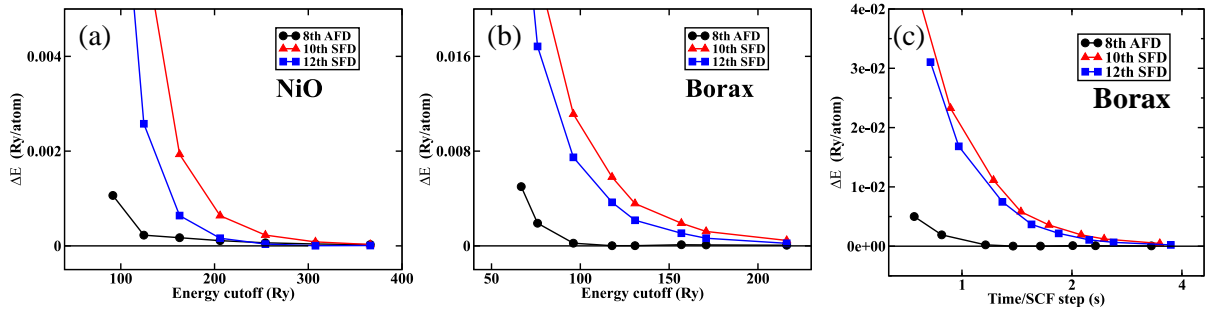

Fig. 4. Differences in total energies between the reference energy obtained using Quantum Espresso and RMG for (a) a 64-atom NiO cell in an antiferromagnetic configuration, and (b) an 86-atom borax decahydrate $Na_2[B_4O_5(OH)_4]\cdot 8H_2O$ cell, as functions of plane-wave kinetic energy cutoff; (c) time per SCF step for grid densities required to reach prescribed accuracies in comparison to the reference energy.

The second test case is borax decahydrate, $Na_2[B_4O_5(OH)_4]\cdot 8H_2O$, which has a complex monoclinic structure with 86 atoms in the primitive cell and a mixture of covalent, ionic, and hydrogen bonds.[86] As shown in Fig. 4(b), the adaptive finite differencing is more accurate than the standard 10th- and 12th-order finite-difference formulas at all tested cutoffs, with the difference exceeding a factor of 200 at 117 Rydbergs, and converges more quickly to the reference value.

Fig. 4(c) shows the time per SCF step at a prescribed accuracy for borax decahydrate, compared to the reference Quantum Espresso plane wave calculation. For the same accuracy, the standard finite difference operators require significantly denser grids than the adaptive operator, and thus their times per SCF step are substantially greater.

**Large-scale DFT calculations using the adaptive operator**

The adaptive finite-difference operator enables high-accuracy calculations for large systems by taking advantage of the enhanced parallelization enabled by domain decomposition when using real-space grids. As an example, we consider a 2,016-atom NiO cell in an antiferromagnetic configuration, also with Ni $3s^2 3p^6$ semi-core states included in valence. Based on the convergence versus grid spacing results in the left panel of Fig. 2, a grid spacing corresponding to an energy cutoff of 167 Ry was selected. The calculation was spin-polarized and used 14,128 Kohn-Sham wavefunctions for each spin channel. The results were obtained using the Frontier development platform Crusher at ORNL. Crusher nodes have the same hardware as the exascale Frontier, but there are only 192 nodes versus Frontier's 9,472. Each node contains a 64-core AMD Epyc 7453 CPU and four AMD MI250X GPUs. The RMG calculations used all CPU cores and all GPUs in each node. The scaling results for 18 to 144 nodes are presented in Fig. 4. A minimum of 18 nodes is required due to the memory footprint of the calculations.

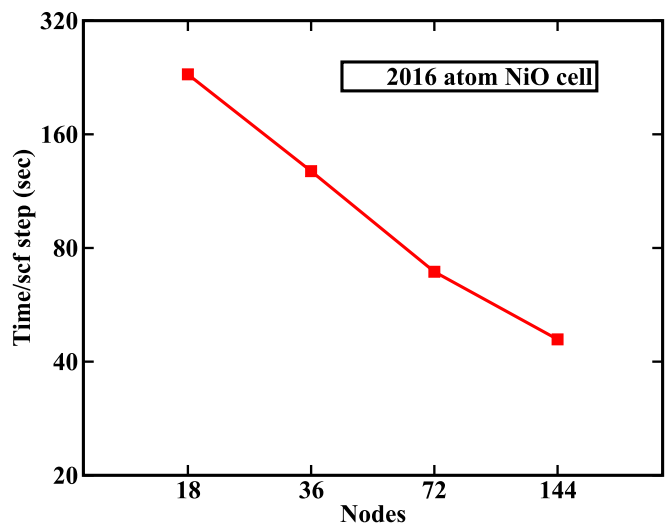

Fig. 5. Scaling of the 2,016-atom NiO calculation with the number of Crusher (Frontier) nodes.

The decrease in the computational time is linear from 18 to 72 nodes, but as the number of nodes increases further, the speedup rate decreases because the computational work per node becomes too small to saturate the nodes fully. It should be noted that even on 18 nodes, the time per SCF step, 230 seconds, makes high-accuracy calculations of this size practical even when the problem requires extensive geometry optimizations, transition path studies, or ab initio molecular dynamics.

**DISCUSSION**

The promise of real-space electronic structure calculations has always centered on their easy parallelizability on massively parallel computer architectures through domain decomposition, because real-space operations are inherently local and thus amenable to parallel execution without extensive communication. However, for grid-based calculations, it is well-established that high-order discretization of the kinetic energy operator is required, which makes the calculations more costly and less local by increasing the communications beyond the local domain. In contrast, plane-wave calculations represent the kinetic energy operator exactly for a given cutoff but require global fast Fourier transform operations. The adaptive finite-differencing scheme developed in this paper relies on optimizing the kinetic energy expression near atomic cores for each atomic specie separately, resulting in finite-difference coefficients that are transferable to different atomic environments in analogy with pseudopotentials, which both grid- and plane-wave-based methods use. The resulting adaptive kinetic energy operator is far more accurate than standard high-order discretizations and lower-order, decreasing the costs of both calculations and communications. We have tested the accuracy of this operator in electronic structure calculations using the well-known Δ test for 71 elemental solids, and the average error of the calculations was the same as those of the well-established plane-wave codes VASP and Quantum Espresso. The accuracy of the adaptive operator was further established with multi-specie tests on NiO and borax decahydrate, which exhibit a range of complex bonding arrangements. The scalability of real-space grid methodology was then confirmed in highly-accurate calculations for a 2,016-atom NiO supercell, where the computational time decreased nearly linearly when scaling from 18 to 144 CPU-GPU nodes.

**METHODS**

The adaptive finite difference operator for the kinetic energy was implemented in version 5 of the open-source RMG code for all crystallographic space groups. The calculations for the Mott insulator NiO used the DFT+U formalism[87] with PBE exchange-correlation functional[88] and U = 6.5 eV. Due to supercell sizes, only Γ point sampling was used. The calculations for borax decahydrate used the PBE functional and were also at Γ. The Δ test calculations used structures provided at https://molmod.ugent.be/sites/default/files/Delta_v3-1_0.zip with varying k-point meshes depending on the specific element. All input and output data are available from the depository listed below.

## DATA AVAILABILITY

All input and output files used to report the results and create the graphics are available at https://github.com/RMGDFT/AFD_5.3.

## CODE AVAILABILITY

The open-source RMG code is distributed through github.com/RMGDFT.

## Appendix 1. Grid spacings for Delta tests

Table A1. Average grid spacings $h$ for different elements in the Delta tests. For a general lattice type, h is defined as $h = \sqrt[3]{V/(nx*ny*nz)}$, where V is the volume of the unit cell at equilibrium, $nx$, $ny$, and $nz$ are the numbers of grid points along the three lattice vectors.

| Element | h (bohr) | Element | h (bohr) | Element | h (bohr) | Element | h (bohr) |
|---------|----------|---------|----------|---------|----------|---------|----------|
| Ag | 0.248 | Fe | 0.212 | Na | 0.209 | Sc | 0.252 |
| Al | 0.240 | F  | 0.236 | Nb | 0.208 | Se | 0.289 |
| Ar | 0.221 | Ga | 0.246 | Ne | 0.229 | Si | 0.271 |
| As | 0.281 | Ge | 0.245 | Ni | 0.210 | Sn | 0.248 |
| Au | 0.207 | He | 0.211 | N  | 0.244 | Sr | 0.358 |
| Ba | 0.209 | Hf | 0.229 | Os | 0.203 | S  | 0.305 |
| Be | 0.259 | Hg | 0.241 | O  | 0.251 | Ta | 0.208 |
| Bi | 0.248 | H  | 0.255 | Pb | 0.250 | Tc | 0.249 |
| Br | 0.243 | In | 0.307 | Pd | 0.235 | Te | 0.259 |
| B  | 0.262 | Ir | 0.192 | Po | 0.264 | Ti | 0.264 |
| Ca | 0.274 | I  | 0.347 | Pt | 0.197 | Tl | 0.264 |
| Cd | 0.260 | Kr | 0.213 | P  | 0.260 | V  | 0.225 |
| Cl | 0.280 | K  | 0.220 | Rb | 0.213 | W  | 0.299 |
| Co | 0.274 | Li | 0.228 | Re | 0.213 | Xe | 0.233 |
| Cr | 0.226 | Lu | 0.169 | Rh | 0.229 | Y  | 0.268 |
| Cs | 0.210 | Mg | 0.289 | Rn | 0.214 | Zn | 0.235 |
| Cu | 0.180 | Mn | 0.243 | Ru | 0.208 | Zr | 0.248 |
| C  | 0.280 | Mo | 0.238 | Sb | 0.236 |    |       |

**ACKNOWLEDGMENTS**

This research was supported by the U.S. Department of Energy's Exascale Computing Project (ECP), Project Number: 17-SC-20-SC. Computing resources at the Oak Ridge Leadership Computing Facility were provided through the ECP User Program and the Innovative and Novel Computational Impact on Theory and Experiment (INCITE) program.

**AUTHOR CONTRIBUTIONS**

E.B. and W.L. devised the adaptive operator and performed the calculations. All authors analyzed the results and wrote the manuscript.

**COMPETING INTERESTS**

The authors declare no competing interests.